\documentclass[a4paper,11pt]{article}
\usepackage[T2A]{fontenc}    
\usepackage{inputenc}
\inputencoding{cp866}
\usepackage{amsmath,amsthm,amssymb}
\begin{document}	
\title{Phase--sensitive quantum effects\\
in Andreev conductance of the SNS system of metals\\
with macroscopic phase breaking length}
\author{Yu. N. Chiang and O. G. Shevchenko\\[5mm]
B. Verkin Institute for Low Temperature Physics and Engineering,\\
National Academy of Sciences of Ukraine\\ 
Kharkov, Ukraine}
\date{\today}
\maketitle
\renewcommand{\abstractname}{}
\begin{abstract}
The dissipative component of electron transport through the doubly connected 
SNS Andreev interferometer 
indium (S) -- aluminium (N) -- indium (S) has been studied. Within helium 
temperature range, the conductance of the individual sections of the 
interferometer exhibits phase--sensitive oscillations of 
quantum--interference nature. In the non--domain (normal) state of indium 
narrowing adjacent to NS interface, the nonresonance oscillations have been 
observed, with the period inversely proportional to the area of the 
interferometer orifice. In the domain intermediate state of the narrowing, 
the magneto--temperature resistive oscillations appeared, with the period 
determined by the coherence length in the magnetic field equal to the critical 
one. The oscillating component of resonance form has been observed in the 
conductance of the macroscopic N--aluminium part of the system. The phase of 
the oscillations appears to be shifted by $\pi$ compared to that of 
nonresonance oscillations. We offer an explanation in terms of the 
contribution into Josephson current from the coherent quasiparticles with 
energies of order of the Thouless energy. The behavior of dissipative 
transport with temperature has been studied in a clean normal metal in the 
vicinity of a single point NS contact.
\end{abstract}
\section{Introduction}
In our previous experiments [1--4] on {\it SNS} structures based on clean 
metals, it has been established that at not too low helium temperatures, the 
dependence of normal conductance on coherent phase difference between 
superconducting banks can be preserved even in case the separation {\it L} 
between the {\it NS} interfaces exceeds the dimension $L \approx 1\mu$m of 
the normal layers in {\it SNS} disordered nanostructures by three orders of 
magnitude. It is in those nanostructures that quantum--interference phenomena 
in dissipative transport have first been observed and are being widely 
explored today [5--13]. This means that the phase--breaking length 
$L_{\phi}^{p}$ in clean metals exceeds that length in nanostructures, 
$L_{\phi}^{d} \sim 1\mu$m, by no less than the same value. Low $L_{\phi}^{d}$ 
in nanostructures seems to be closely related to the short elastic scattering 
length for electrons: $l_{el} \sim 0.01\mu$m in 3D--structures and $l_{el} 
\lesssim 1\mu$m in 2D--electron gas. The above magnitudes of $l_{el}$ in 
metals are typical of high concentration of lattice defects. Insignificant 
contribution from inelastic scattering at the objects of that kind is 
evidently the main factor which constrains $L_{\phi}$. In contrast, the 
macroscopic value of $L_{\phi}$ in clean metals enabled us to extend the 
spatial range for examining the phase--coherent phenomena and to succeed in 
observing for the first time the long--range phase coherence at the ratios 
$L/\xi_{T}>1$, up to $10^{2}$.

Moreover, note that the coherence length $\xi_{T}^{d}=\sqrt{\hbar D/k_{\rm B}T}\ 
(D$ is the diffusion constant) in dirty (diffusive) limit is expressible in 
terms of that length $\xi_{T}^{p} \equiv \xi_{T}^{bal}=\hbar v_{\rm F}/
k_{\rm B}T$ in clean limit as follows
$$\xi_{T}^{d}=[(1/3)l_{el} \xi_{T}^{bal}]^{1/2}, \qquad l_{el} \ll 
\xi_{T}^{bal}.$$
Consequently, the respective temperature ranges $T^{p}$ and $T^{d}$ for 
clean and dirty specimens in which the values $L/\xi_{T}$ are the same, do not 
coincide but must be related by the equation
\begin{equation}\label{1}
\frac{(T^{p})^{2}}{T^{d}}=3\frac{\hbar v_{\rm F}}{k_{\rm B}l_{el}^{d}}
\left( \frac{L^{d}}{L^{p}} \right)[\frac{(L/\xi_{T})^{p}}
{(L/\xi_{T})^{d}}]^{2}.
\end{equation}
(We imply that the phase--breaking length $L_{\phi}$ is no less than both 
$L^{p}$ and $L^{d}$, the separation between {\it NS} interfaces in clean and 
dirty samples, respectively). It thus follows from Eq. (1) that the 
values of the parameter $L/\xi_{T}=\sqrt{T/E_{c}}\ (E_{c}$ is the gap in the 
density of states [14]) common to both limits at which phase--coherent 
phenomena behave similarly must be realized at different temperatures. For 
clean metals, these would be significantly higher. For example, for 
$L^{p}/L^{d} \sim 10$, the parameter $L/\xi_{T}$ in a clean sample with 
$l_{el} \gg 1\mu$m at $T=2$K is of the same order of magnitude as that in a 
diffusive sample with $l_{el} \sim 0.01\mu$m at $T \lesssim 0.1$K. Below we 
will show that the relationship between the temperature regions within which 
phase--coherent effects behave analogously in 2DEG--samples with $l_{el} \sim 
1\mu$m [11] and 3D--samples with $l_{el} \sim 0.01\mu$m [5--7], provided 
$L^{3D}/L^{2D} \sim 1$, is also given by Eq. (1).

The initiation of the phase--coherent phenomena at $L/\xi_{T} >1$ means that 
long--range phase coherence exists under exponentially small proximity effect 
for the main group of quasiparticles with the energies $\epsilon \sim T$. In 
ultra--clean structures those phenomena can therefore be observed within a 
macroscopic scale and at not too low helium temperatures. This circumstance 
may appear to be urgent when solving the problem of extracting certain 
quantum information from various quantum systems through macroscopic channels.

The first evidence for the long--range influence of a superconductor on the 
conductivity of a normal metal adjacent to it is directly contained in the 
experimental observables from the structures with a single {\it NS} boundary 
[15, 16]. The effect was reported to extend over length scales up to 
$L/\xi_{T} \sim 5 \div 10$ away from the boundary. Subsequently, the 
interference effects have been discovered in doubly connected {\it SNS} 
systems made of disordered metals (nanostructures) with short $L_{\phi}$ 
[5--7]. Up to now, the variety of $L/\xi_{T}$ studied did not exceed in 
magnitude the abovementioned. However, as our experiments show, the 
manifestation of the phase--coherent phenomena in doubly connected {\it SNS} 
systems is not restricted by that interval of $L/\xi_{T}$. Besides, 
available findings concerning the behavior of those phenomena within the 
limits of that interval are sometimes interpreted ambiguously (see Fig. 6). 
Thus, further investigations of the phenomena of such kind are needed.

Below we report on the investigations of the temperature and phase--sensitive 
features of the conductance of the {\it SNS} system in a geometry of an 
Andreev interferometer. The system was formed by two clean metals in contact, 
aluminium (in the normal state) and indium, both with $l_{el} \approx 100
\mu$m. This allows us to achieve the conditions $L,\ l_{el} \gg \xi_{T}=
\xi_{T}^{p}$. The ratio $L/\xi_{T}$ was about $10^{2}$. The contribution from 
supercurrent due to the main group of carriers with the energy 
$\epsilon \sim T$ was entirely eliminated since all three dimensions of the 
normal layer in the {\it SNS} system exceeded inherent microscopic spatial 
parameters which are responsible for the proximity effect.
\section{Experiment}
Shown in Fig. 1 is a schematic picture of the doubly connected system 
investigated made of two metals in contact, aluminium and indium; Inset is an 
equivalent measuring scheme. Once indium becomes superconducting, the system 
acquires the {\it SNS} configuration of Andreev--interferometer type with an 
orifice formed by an aluminium bar (of cross section $2 \times 2$mm) and 
an indium strip soldered to each other at the points {\it a} and {\it b}. 
The orifice area comprises $A=ab \times h \approx 3\mbox{mm} \times 15\mu$m. 

In our early experiments involving the {\it SNS} systems with copper [3, 4], 
the wide soldered {\it NS} contacts of characteristic size $\overline m$ 
could not considerably increase the contact resistance $R_{cont}$ since 
${\overline m} \gg l_{el}$. In contrast, here, current flows through the 
narrow contacts {\it a} and {\it b} with significant spreaded resistance 
$R_{\rm Sh}$ of "Sharvin type" which usually develops as ${\overline m} \ll 
l_{el}$ [17]. The contacts of such size appear in spot soldering indium to 
aluminium we use in the present work. Note that as we confirmed repeatedly 
before, the immediate soldering of the metals of the highest purity, with 
Residual Resistance Ratio at 300K and 4.2K $RRR \gtrsim 10^{4}\ (l_{el} 
\approx 100\mu$m), results in the contact barrier height {\it z} close to zero, 
the corresponding transparency coefficient being $t=(1+z^{2})^{-1} \approx 
1$ [18]. ($z \neq 0$ for processing that do not destruct, fully or partially, 
an oxide layer or other contaminations at the metallic surface).

We can estimate the characteristic dimensions of the contacts {\it a} and 
{\it b} if note that at $l_{el}> \overline m$, the total current $I_{NN}$ 
through a contact between two normal metals must be related to the contact 
area $A_{cont}$ by the equation [18]
\begin{equation}\label{2}
I_{NN}=2 \nu(\epsilon)e^{2}v_{\rm F}A_{cont}U_{\rm Sh}t \equiv U_{\rm Sh}/
R_{\rm Sh}
\end{equation}
in which $\nu(\epsilon)$ is the density of states in either contacted metal 
while $U_{\rm Sh}$ is the voltage drop at the spreaded resistance 
$R_{\rm Sh}$. Selecting aluminium, with normal conductivity $\sigma_{N}=
(1/3)e^{2}v_{\rm F}\nu(\epsilon)l_{el}$, as a 3D--part of the system, from Eq. 
(2) we get the contact area $A_{cont}$
\begin{equation}\label{3}
A_{cont}=(1/6)(l_{el}^{\rm Al}/L^{\rm Al})(U^{\rm Al}/U_{\rm Sh})A^{\rm Al}.
\end{equation}
Here, we took into consideration that $I_{NN}=j^{\rm Al}A^{\rm Al}=j_{cont}
A_{cont}\ (j^{\rm Al}$ and $j_{cont}$ are the current densities in aluminium 
and in the contact, respectively). In addition, $A^{\rm Al} \approx 4{\rm mm}
^{2}$ is the cross section area of aluminium bar; $L^{\rm Al} \approx 1.5$mm 
is the length of the corresponding Al part between one of the contacts, say 
{\it a}, and the measuring probe $V_{2}$ ({\it ac} in Fig. 1); $U^{\rm Al}=
I_{NN}R^{\rm Al}$ is the voltage across the Al part; $R^{\rm Al}$ is the 
resistance of that part measured independently. The potential difference 
$U_{\rm Sh}$ across the spreaded resistance can be found from the voltage 
{\it U} measured at the probes $V_{1}$ and $V_{2}$:
$$U_{\rm Sh}=U-I_{NN}(R^{\rm Al}+R_{narr,N}^{\rm In})$$
where $R_{narr,N}^{\rm In}$ is the resistance of the indium narrowing next to 
the contact (see Inset to Fig. 1) in the normal state. Other 
quantities necessary for estimating $A_{cont}$ have been measured to be as 
follows. 
$$R^{\rm Al} \approx 4 \cdot 10^{-10} \Omega;\ R_{\rm Sh} \approx 1.1 \cdot 10^{-8}
\Omega;\ R_{narr,N}^{\rm In} \approx 1.7 \cdot 10^{-8}\Omega.$$ 
Hence, Eq. (3) yields that the characteristic dimension $\overline m$ 
of the "spot" at the contacts {\it a} and {\it b} may amount to approximately 
25$\mu$m, this corresponding to inequality $l_{el}> {\overline m}$, so that 
an additional spreaded resistance $R_{\rm Sh}$ may appear. In our experiment, 
$R_{\rm Sh}$ exceeds the resistance of the normal region {\it ac} by two 
orders of magnitude.

We performed dc four-- and three--terminal measurements using normal (copper) 
leads $I_{1},\ V_{1}$ and $I_{2},\ V_{2}$. The former were soldered to indium 
beyond the narrowings while the latter were spot welded onto aluminium. 
Measuring current ($I \approx 0.5$A) was inserted into the system via the 
leads $I_{1},\ I_{2}$. Once the {\it NS} contacts have been prepared, the 
indium narrowing next to the contact {\it b} was further thinned down (by 
drawing) to bring the resistances of the interferometer arms {\it dbf} and 
{\it daf} (see Inset to Fig. 1) into the relation $R_{dbf} \gg R_{daf}\ 
(R_{dbf} \equiv R_{narr,b}^{\rm In} \sim 10^{-3}\Omega)$. Assuming this, 
practically all the current injected was passed through the circuit 
"$I_{1}$ -- indium narrowing -- contact {\it a} -- aluminium -- $I_{2}$". 
The macroscopic phase difference was still controllable.

The phase difference was varied by applying external magnetic field $H_{e}$ 
from the rectangular wire turn carrying the current $I_{H_{e}}$. The turn was 
attached directly onto the aluminium bar face in such a way that the plane of 
the interferometer orifice was parallel to that of the turn. This circumstance 
simplifies calculating the field strength in the orifice region. To compensate 
the extraneous fields, the sample with the turn was placed inside a closed 
superconducting screen.

The potential difference between leads $V_{1},\ V_{2}$ was measured by the 
device using a thermomagnetic superconducting modulator [19], with an accuracy 
to no more than $(0.5 \div 1) \times 10^{-12}$V. This allows us to study the 
effects of the magnitude of down to 0.1\% in the conductance of macroscopic 
{\it N}--regions. The error in measuring current and temperature ranges from 0.001\% 
up to 0.01\%. Current--voltage characteristics were verified to be linear 
over a wide interval of currents.
\section{Results and discussion}
\subsection{$H_{e}=0$. Temperature dependence}
Curves 1 and 2 in Fig. 2 depict the temperature dependence of the potential 
difference {\it U} normalized by the measuring current which is inserted into 
the system via the leads $I_{1}$ and $I_{2}$, assuming $R_{cont}^{a} \ll 
R_{cont}^{b}$. When the temperature was lowered down to the critical 
superconducting point for the bulk indium part $T_{c}^{\rm In}=3.41$K 
and the {\it NS} boundary developed, a step--like increase in the resistance 
of {\it eac}--section was revealed, of the type we pioneered in observing in 
1988 [15]. We believe this to be a characteristic quantum effect which 
accompanies the initiation of Andreev reflection [20]. Note that Andreev 
reflection can manifest itself macroscopically only at temperatures somewhat 
below $T_{c}$ for bulk indium when the superconducting energy gap grows in 
value noticeably. In particular, at $T=3.2{\rm K},\ \Delta(T) \approx 0.1 
\Delta(0)\ \mbox{while}\ [1-T/T_{c}] \approx 0.06$. Analyzing the contribution 
into resistance from individual parts of the {\it SNS} system, estimated in 
Sec. 2, and curves 1 and 2 in Fig. 2, one concludes that only indium narrowing 
can be responsible for the height of the jump in resistance near 
$T_{c}^{\rm In}$ and its further changing with cooling down to 
$T \approx 1.8$K ($R_{\rm Sh}$ does not depend on temperature and 
$R^{\rm Al} \ll R_{narr,N}^{\rm In}$). It is seen that the resistance of the 
narrowing at {\it NS} configuration of the system ($\approx 3.4 
\times 10^{-8}\Omega\ \mbox{at}\ T=3.2$K) is twice as large as that 
resistance in {\it NN} state ($\approx 1.7 \times 10^{-8} \Omega\ \mbox{at}\ 
T=3.5$K).

According to the microscopic theory [21, 22] such an increase in the normal 
resistance assuming Andreev reflection occurs is due to the doubling of the 
scattering cross section for electrons at the impurities located within the 
range of order of the coherence length $\xi_{T}^{p}$ away from the {\it NS} 
boundary (for indium, $\xi_{T}^{p} \approx 10\mu$m at $T \approx 3$K). This 
can be detected in case $L \sim \xi_{T}^{p}$ where {\it L} is the length of 
a normal metal layer measured from the boundary. The estimation for the 
dimensions of the narrowing, with the magnitudes of ${\rm RRR^{\rm In}} 
\approx 4 \times 10^{4},\ A_{cont},\ \mbox{and}\ R_{narr}^{\rm In}$, shows 
that the distance from the "spot" to the bulk indium section of the system 
where the {\it NS} boundary arises at $T < T_{c}^{\rm In}$, is of order of 
$10\mu$m, i. e., is comparable to $\xi_{T}^{p}$. Therefore, the above 
theoretical conclusion about the twofold enhancement of the resistance 
seems to be directly confirmed for the first time. The maximum increase in 
the resistance we managed to observe before did not exceed 60\% [23].

In Fig. 3 (curves 2--5) we present the conductance measured on the opposite 
side from the contact {\it a}, within the normal aluminium part, as a function 
of the thickness of the normal layer next to {\it NS} boundary, i. e., of the 
separation $L_{NS}$ between the normal lead {\it N} and the superconducting 
point contact {\it a}. The measurements were performed by dc four--terminal 
zero method which allowed us to exclude the contribution $R_{\rm Sh}+
R_{narr}^{\rm In}$. In this case, the ring of the interferometer was 
interrupted. For comparison, in Fig. 3, we also show the 
temperature--dependent resistance of the same aluminium sample (curve 1) 
measured using only normal leads.

The curves in Fig. 3 illustrate the evolution of the increase in resistance 
of the next--to--contact aluminium layer dependent on $L_{NS}$, with the 
rise of {\it NS} boundary. It is seen from comparing Figs. 2 and 3 that the 
effect of increasing in the normal resistance observed on each side of the 
contact {\it a} is similar to the effects evidenced in other {\it NS} 
systems, with other metals, at arbitrary area of the {\it NS} boundaries, 
and dissimilar arrangements of the leads [2, 23]. The nature of the effect as 
mentioned above is associated with the interference of the coherent Andreev 
reflected electrons while its magnitude only depends on the ratios 
$\xi_{T}/L_{NS},\ l_{el}$ provided $L_{\phi} \gg l_{el} \gg \xi_{T}\ 
\mbox{and}\ L_{NS} < L_{\phi}$. The results in Fig. 3 thus show once again 
that the long--range phase coherence in a clean metal at the temperatures 
investigated can be sustained within macroscopic distances, of no less than 
1.5mm in our case, at $L/\xi_{T} \approx 10^{2}$. This fact, as well as our 
previous findings [1--4, 23], points out that the phase breaking length is 
at least of the same order or greater.

The temperature--dependent resistance of both indium and aluminium measured 
on each side of the contact {\it a} below the jump temperature where 
$\xi_{T} < L_{NS}$, is governed by the same power law $\sim T^{3.5}$ (see 
Figs. 2 and 3). In Ref. 2, we detected similar behavior of aluminium 
comprising a part of {\it NS} system when measurements were carried out in 
a different way. We find this to be an additional confirmation that the 
temperature--dependent phase--breaking length does determine the temperature 
dependence of the conductance of a metal layer as a whole, within the range 
$\xi_{T} < L_{NS} <L_{\phi}$, under multiple Andreev reflections [2].
\subsection{$H_{e} \ne 0$}
\subsubsection{Nonresonance oscillations}
The potential difference {\it U} measured across the leads $V_{1},\ V_{2}$ 
as a function of the external magnetic field $H_{e}\ \mbox{at}\ T=3.2$K 
exhibits an oscillating component with the period $(hc/2e)/A,\ A$ being the 
area of the orifice (see Fig. 1). The amplitude of the oscillations is 
plotted in Fig. 4, curve 1, in relative units $U/I \propto (R_{H}-
R_{H=0})/R_{H=0}$. Its absolute value comprises $\Delta (U/I)=(R_{max}-
R_{min}) \approx 4.5 \times 10^{-10}\Omega$ which corresponds to 
approximately 2\% of the indium narrowing resistance $R_{narr,SN}^{\rm In}$. 
Here, $R_{min}\ \mbox{stands for}\ R_{H=0}$. In Fig. 5, the temperature 
dependence of the difference $\Delta (U/I)=(R_{H=0.3{\rm mOe}}-R_{H=0})$ is 
displayed. The position of the step on this curve, along with that on the 
dependence ${\rm d}(U/I)/{\rm d}T$ in Fig. 2, points out that the domain 
intermediate state in indium narrowing is realized only after reducing the 
temperature down to $T \approx 3.1$K. Independent analysis, accounting for 
the size $L_{narr}^{In}$, leads to the same conclusion. Indeed, at 
$T \gtrsim 3$K, the length $L_{narr}$ in self--magnetic field ($\sim 10$Oe) 
of measuring current does not satisfy the condition for arising the domain 
structure with the number of domains greater than 1 [3]. Moreover, once the 
temperature reduced lower than 3K, magneto--temperature resistive 
oscillations appear (see Inset to Fig. 5), with the period $\Delta_{H_{c}(T)} 
\sim hc/e\xi_{H}^{2}$ in critical magnetic field where $\xi_{H} \approx 2 
\sqrt{qR_{\rm L}[H_{c}(T)]} \sim 1\mu \mbox{m\ at}\ T=3.0$K ({\it q} is the 
screening radius for an impurity, $R_{\rm L}(H_{c})$ the Larmour radius) 
[1, 3]. This fact is supposed to result from the transition of the indium 
narrowing into the domain intermediate state and thus is an additional 
evidence for that transition occurs at temperatures not higher than 
$\approx 3.1$K.

Compare the parameters of oscillations observed at 3.2K (curve 1 in Fig. 4) 
with theory and the data thus far available from other investigators. In 
Fig. 6, we plotted the most characteristic data from Refs. 7, 11, and 13 on 
the temperature dependences of the relative amplitudes $\vert \Delta R/R_{N} 
\vert$ of the resistive oscillations as a function of the parameter 
$T_{\rm Th}/T \equiv (\xi_{T}/L)^{2}$, with the "Thouless temperatures" 
$T_{\rm Th}$ adopted by the authors. Also, shown is the theoretical curve [24] 
$\vert \Delta R/R_{N} \vert =\vert R_{max}-R_{N} \vert /R_{N},\ \mbox{where}\ 
R_{max}\ \mbox{and}\ R_{N}$ are the resistances in maximum and minimum of 
the oscillations, respectively. The curve has been received by numerical 
simulation for cases $\phi=\pi\ \mbox{and}\ \phi=0$ assuming $T_{\rm Th}=
D/\pi L^{2}$. Apperent discrepancy between the data presented and theory [24], 
as well as between different experiments, can be almost entirely removed if 
one takes an energy criterium $T^{*}=D/2\pi L^{2}$ received for dirty limit 
in Ref. 25 as a gap in density of states generated when coherent excitations 
are localized in a normal part between {\it NS} interfaces due to Andreev 
reflections. The same results as in Fig. 6 are presented in Fig. 7b, using 
the separation between superconducting "mirrors" as {\it L} and the above 
parameter $T^{*}$ as "Thouless temperature". It can be seen that the 
experimental oscillation amplitudes modified in such a way follow a certain 
law in the parameter $T^{*}/T$. This feature results immediately from the 
theory by Aslamazov, Larkin, and Ovchinnikov [25] developed as early as 1968. 
In fact, consider that quasiparticle dissipative current is a difference 
between the total current and its non--dissipative part and is proportional 
to $f(\cos \Delta \chi),\ \Delta \chi$ being the macroscopic phase difference 
[26]. Based on the analytical expressions for non--dissipative current 
($\sim \sin \Delta \chi$) from [25] we thus find
\begin{equation}\label{4}
\frac{R_{\Delta \chi=\pi/2}}{R_{N}}=[1-(1/\pi) \frac{L}{\xi_{T}} \exp 
(-(\frac{L}{\xi_{T}}+1)) \ln (\alpha (\frac{L}{\xi_{T}})^{-2})]^{-1}.
\end{equation}
Here, $L/\xi_{T} \equiv (T/T^{*})^{1/2},\ \alpha$ is a coefficient of order 
of unity.

The curve thus calculated for $\alpha=2$ is plotted in Fig. 7a, together with 
the curve $R_{\Delta \chi=0}/R_{N}$ from Ref. 24. It can easily be seen that 
both curves predict the existence of long--range phase coherence, i. e., a 
non--exponential decay of the oscillating dissipative component in the 
conductance of an {\it SNS} system under $L/\xi_{T} \gg 1$. The two curves 
differ from one another by the factor $\sqrt{2}$ in their position relative 
to $L/\xi_{T}$ scale to the extent as $\xi_{T}^{[24]}$ differs from 
$\xi_{T}^{[25]}$. The same theoretical curve for the relative oscillation 
amplitude $\vert \Delta R/R_{N} \vert$ from Ref. 25 is shown in Fig. 7b by 
dashed line. It describes properly the position of all the experimental data 
from Fig. 6 on the temperature scale which fact supports the conclusion about 
the relationship between the temperature intervals deduced from Eq. (1). 
(When handling the experimental data from Ref. 11 we took into consideration 
that the normal--part size in one of the directions exceeded $l_{el}$ and did 
not satisfy the ballistic criterion for $\xi_{T}^{p}$. In this case, the 
Thouless temperature must be estimated in a different way as we have done). 
Moreover, from the curve [25], we can also obtain correct quantitative 
estimation for the oscillation amplitude in the corresponding temperature 
intervals. Excluded are the data reported in Ref. 11 where the total sample 
resistance is taken as $R_{N}$ rather than the resistance of the section 
between the "mirrors".

Hence, as the above analysis corroborates, the experimental results 
[7, 11, 13] have most likely been received in the range of the parameter 
$L/\xi_{T} > 1\ (T^{*}/T < 0.3)$, i. e., in "dirty" limit. In contrast, it 
seems to be reasonable to attribute the oscillations observed from the 
indium narrowing next to contact {\it a} at $T=3.2$K to the quasiballistic 
regime $L_{narr}^{\rm In}/\xi_{T} \sim 1$. In this regime, a characteristic 
temperature must be $T^{bal}=\hbar v_{\rm F}/k_{\rm B}L$. The oscillation 
amplitude calculated employing this parameter as $T^{*}$ and $L_{narr}^
{\rm In}\ \mbox{as}\ L$ is shown in Fig. 7b as a square. Its location on the 
temperature scale agrees with theory [25].
\subsubsection{Resonant oscillations}
At $T \sim 2$K, we observe the oscillations in a magnetic field (curve 2 in 
Fig. 4) which have a resonant form unlike those observed at 3.2K. Their 
period does not change and is given by $(hc/2e)/A$. We assume their nature 
to be connected with the peculiarities of the phase--coherent interference 
in aluminium. The reasonings are as follows. First, at $T \sim 2$K, the 
resistance of indium part becomes as low as that of aluminium between points 
{\it a} and {\it c} (see Inset to Fig. 1). Second, the phase of resonant 
oscillations is shifted by $\pi$ relative to that of nonresonance 
oscillations. (It is worth noting that the above inversion of the resistive 
oscillation phase has also been observed in other works, for example, [11, 13] 
in which the interferometer geometry and measuring technique differ from 
those in our experiment).

We should emphasize once more that in the system investigated the 
phase--breaking length is either much greater than the separation between the 
injectors of electrons, as in case of indium narrowing in the domain state, 
or of order of that separation, as in case of aluminium part ($\sim 1$mm in 
length). This is the first condition necessary for phase--coherent 
quasiparticle phenomena to reveal themselves in the conductance of {\it SNS} 
systems with large separation between the interfaces. Next principal 
consideration which has been discussed in detail in theory [27] is the 
limitation on the dimensions of injectors which act as reservoirs of 
quasiparticles. In the ballistic regime, an electron beam must be splitted 
at the injector site in order that Andreev--reflected excitations, with low 
energies $\epsilon \leq E_{c} \sim \hbar v_{\rm F}/L$, follow quasiclassical 
paths connecting both superconducting "mirrors", instead of returning into 
the injector after the first reflection. Under such conditions, the 
coherent phase difference between the "mirrors" can be established. As shown 
[27], the diameter of the injector--reservoir should not exceed de Broglie 
wave length $\lambda_{\rm B}$ (this was first noted in Ref. 28). It is not 
difficult to understand that this limitation loses its meaning if 
a superconducting bank serves as at least one of the injectors since in this 
case the splitting is not needed for initiating a trajectory connecting both 
banks. In our {\it SNS} system, current is introduced through one of the 
"mirrors" (Fig. 1) so that the above limitation is absent, for both the 
indium narrowing in the ballistic regime and the aluminium section in the 
regime close to the diffusive one.

Since the separation between Andreev levels is $\sim \hbar v_{\rm F}/L$ we 
can assume that it is the aluminium part in which 2K--oscillations arise 
connected with the fine structure of Andreev spectrum for low--energy 
electrons. As mentioned, the oscillations have a resonant form, in contrast 
to both 3.2K--oscillations in indium narrowing and the oscillations we 
observed before in normal copper part of the {\it SNS} system with large area 
of "mirrors" and current injected not through the "mirrors" [4]. As can be 
seen from Fig. 4, the amplitude of the resonant oscillations relative to the 
resistance of aluminium between the "mirrors" is about 4\%, in accordance 
with the ratio $E_{c}L/T$ for aluminium rather than indium narrowing. Theory 
[27, 29] yields that the resonant oscillations can be expected to result from 
the degeneracy of the transverse modes on the Fermi level. In this case, the 
Andreev--level energies 
$$\epsilon_{n}^{\pm}=\frac{\hbar v_{\rm F}}{2L}[(2n+1)\pi \mp \Delta \chi]$$
($\Delta \chi$ is the macroscopic coherent phase difference between the 
"mirrors") go to zero as soon as $\Delta \chi=(2n+1)\pi$. The degeneration 
condition therefore assumes that the phase of resonant oscillations should be 
inversed relative to that of nonresonance oscillations, the latter being 
given by $\Delta \chi=2n\pi$. The inversion of this kind we observe for the 
oscillations of resonant form.
\section{Conclusions}
The phase--coherent component of the dissipative electron transport has been 
studied in a doubly connected hybrid system formed by clean metals, In and Al, 
with elastic mean free path about $100\mu$m and phase--breaking length 
greater than 1mm. The device has a geometry of an Andreev {\it SNS} 
interferometer. The characteristic dimension of the {\it NS} interfaces is 
less than the mean free path while the size of the normal part between the 
interfaces is comparable to the macroscopic phase--breaking length. A number 
of phase--sensitive effects of quantum--interference nature have been 
revealed in cases when current is injected both through one of the {\it NS} 
interfaces and beyond them. The effects result from the presence of coherent 
component due to Andreev reflection. We distinguished the effects originating 
from different regions of the {\it SNS} system, indium narrowing in the 
vicinity of the point--contact {\it NS} interface and normal aluminium. 

The resistive oscillations with the period $\Phi_{0}/A$ observed in an 
external magnetic field at $T=3.2$K we relate to the behavior of the electron 
transport in the indium narrowing in the normal (non--domain) state. In the 
domain intermediate state of the narrowing, the oscillations of the 
magneto--temperature type have been revealed, their period being $2\Phi_{0}/
\xi_{H_{c}(T)}$.

At $T \lesssim 2$K, the resistive oscillations of resonant form are detected, 
with the phase shifted by $\pi$ in reference to that of nonresonance 
oscillations. We suggest that the resonant oscillations are exhibited by 
macroscopic normal--aluminium section of the system. The oscillations 
originate from the degeneracy of the transverse Andreev modes for coherent 
quasiparticles, with energies of order of the Thouless energy, at the Fermi 
level. For such quasiparticles, the transport regime may be ballistic when 
moving through a normal region of macroscopic size {\it L}, between a 
reservoir and {\it NS} interfaces, if $l_{el} \gg \xi_{T}^{bal}$. In the 
ballistic regime, on condition that {\it NS} interface serves as one of the 
electron injectors, the manifestation of the phase coherence does not depend 
on {\it L} as long as $L \leq L_{\phi}$. It can thus be assumed that the 
observation of the phase--sensitive effects in the conductance of 
macroscopic {\it SNS} systems is only restricted by those values of {\it L} 
at which the normal reflection from {\it NS} interfaces becomes dominant. 
This occurs provided $E_{c}/T < \sqrt{E_{c}/E_{\rm F}}$ [27] whence it 
follows that at $T \sim 2{\rm K}\ \mbox{and}\ L_{\phi} \rightarrow \infty\ L$ 
must be over 10cm. This is a limiting value of the distance between 
interfaces at which the long--range phase coherence can manifest itself at 
helium temperatures ($(L/\xi_{T}^{bal}) \sim 10^{4}$).

The phase--sensitive quantum phenomena in the conductance of the {\it SNS} 
system formed by clean metals we observed experimentally at not too low helium 
temperatures under the condition $(L/\xi_{T}^{bal}) \sim 10^{2}$ can be 
reasonably explained in the limits of the above {\it L}--scale for 
long--range phase coherence due to the contribution from low--energy coherent 
excitations with the energy $E_{c} \ll T,\ \Delta$.

\newpage
\begin{center}
FIGURE CAPTIONS
\end{center}
Fig. 1. Sketch of the {\it SNS} interferometer and equivalent 
measuring scheme (Inset). Crosshatched is the bulk part of indium.

\vspace{0.5cm}

Fig. 2. Temperature--dependent resistance of indium narrowing in the vicinity 
of contact {\em a} at $R^{a}_{cont} \ll R^{b}_{cont}$ (curves 1, 2) and its 
derivative (curve 3).
Circles and triangles depict the data from minimum ($H_{e}=0$) and maximum 
($H_{e}=0.3$mOe) of the resistive oscillations observed at $T=3.2$K.
The jump on the curves 1, 2 corresponds to the twofold increase in the 
resistance of indium narrowing after the initiation of the {\it NS} interface 
(spreaded contact resistance included).

\vspace{0.5cm}

Fig. 3. Effect of increasing in the resistance of aluminium part near the 
interface (in the region {\it ac}, see Fig. 1) as a function of the 
separation between {\it N} probe and point {\it NS} interface. $H_{e}=0$.

\vspace{0.5cm}

Fig. 4. Phase--sensitive component of the resistance of the 
interferometer with $R_{a} \ll R_{b}$ vs external magnetic field. 
Nonresonance oscillations from indium narrowing at $T=3.2$K (triangles) and 
resonant oscillations from aluminium at $T=2$K (circles).

\vspace{0.5cm}

Fig. 5. Temperature dependence of the difference between the resistance of 
the interferometer at $H_{e}=0.3\ \mbox{mOe\ and}\ H_{e}=0$ (Inset is the 
same enlarged).

\vspace{0.5cm}

Fig. 6. Temperature dependence of the amplitude of phase--sensitive 
nonresonance conductance oscillations from experiments [7], [11], and [13]. 
Solid curve is theory [24].

\vspace{0.5cm}

Fig. 7. {\it a} -- Resistance of an {\it SNS} system against the 
parameter $L/\xi_{\rm T}$ as follows from [24] and [25]; 
{\it b} -- Temperature dependence of the amplitude of nonresonance 
conductance oscillations from [7], [11], and [13] (see Fig. 7) modified in 
accordance with the theory [25].

\begin{thebibliography}{99}
\bibitem{1}Yu. N. Chiang (Tsyan), Pis'ma Zh. Eksp. Teor. Fiz. {\bf 71}, 
481 (2000) [JETP Lett. {\bf 71}, 334 (2000)].
\bibitem{2}Yu. N. Chiang, S. N. Gritsenko, and O. G. Shevchenko, Zh. Eksp. 
Teor. Fiz. {\bf 118}, 1426 (2000) [JETP {\bf 91}, 1235 (2000)].
\bibitem{3}Yu. N. Chiang and O. G. Shevchenko, Fiz. Nizk. Temp. {\bf 27}, 
1357 (2001) [Low Temp. Phys. {\bf 27}, 1000 (2001)].
\bibitem{4}Yu. N. Chiang and O. G. Shevchenko, Pis'ma Zh. Eksp. Teor. Fiz. 
{\bf 76}, 794 (2002) [JETP Lett. {\bf 76}, 670 (2002)].
\bibitem{5}V. T. Petrashov, V. N. Antonov, P. Delsing, and T. Claeson, Phys. 
Rev. Lett. {\bf 70}, 347 (1993); {\bf 74}, 5268 (1995).
\bibitem{6}A. Dimoulas, J. P. Heida, B. J. v. Wees, T. M. Klapwijk, W. v. d. 
Graaf, and G. Borghs, Phys. Rev. Lett. {\bf 74}, 602 (1995).
\bibitem{7}H. Courtois, Ph. Gandit, D. Mailly, and B. Pannetier, Phys. Rev. 
Lett. {\bf 76}, 130 (1996).
\bibitem{8}P. Charlat, H. Courtois, Ph. Gandit, D. Mailly, A. F. Volkov, and 
B. Pannetier, Phys. Rev. Lett. {\bf 77}, 4950 (1996).
\bibitem{9}H. Takayanagi and T. Akazaki, Phys. Rev. B {\bf 52}, R8633 
(1995); Physica B {\bf 249--251}, 462 (1998).
\bibitem{10}S. G. den Hartog, B. J. van Wees, Yu. V. Nazarov, T. M. Klapwijk, 
and G. Borghs, Physica B {\bf 249--251}, 467 (1998).
\bibitem{11}E. Toyoda and H. Takayanagi, Physica B {\bf 249--251}, 472 (1998).
\bibitem{12}P. G. N. de Vegvar, T. A. Fulton, W. H. Mallison, and R. E. 
Miller, Phys. Rev. Lett {\bf 73}, 1416 (1994).
\bibitem{13}A. Kadigrobov, L. Y. Gorelik, R. I. Shekhter, M. Jonson, R. Sh.
Shaikhaidarov, V. T. Petrashov, P. Delsing, and T. Claeson, Phys. Rev. B 
{\bf 60},14589 (1999).
\bibitem{14}F. Zhou, P. Charlat, B. Spivak, and B. Pannetier, J. Low Temp. 
Phys. {\bf 110}, 841 (1998).
\bibitem{15}Yu. N. Chiang (Tszyan) and O. G. Shevchenko, Fiz. Nizk. Temp. 
{\bf 14}, 543 (1988) [Sov. J. Low Temp. Phys. {\bf 14}, 299 (1988)].
\bibitem{16}V. T. Petrashov, V. N. Antonov, S. V. Maksimov, and 
R. Sh. Shaikhaidarov, Pis'ma Zh. Eksp. Teor. Fiz. {\bf 58}, 48 (1993) [JETP 
Lett. {\bf 58}, 49 (1993)].
\bibitem{17}Yu. V. Sharvin, Zh. Eksp. Teor. Fiz. {\bf 48}, 984 (1965).
\bibitem{18}G. E. Blonder, M. Tinkham, and T. M. Klapwijk, Phys. Rev. B
{\bf 25}, 4515 (1982).
\bibitem{19}Yu. N. Chiang, Prib. Tekhn. Eksp. {\bf 1}, 202 (1981).
\bibitem{20}A. F. Andreev, Zh. Eksp. Teor. Fiz. {\bf 46}, 1823 (1964) [Sov. 
Phys. -- JETP {\bf 19}, 1228 (1964)]; {\bf 49}, 655 (1965) [{\bf 22}, 455 
(1965)].
\bibitem{21}J. Herath and D. Rainer, Physica C {\bf 161}, 209 (1989).
\bibitem{22}A. M. Kadigrobov, Fiz. Nizk. Temp. {\bf 19}, 943 (1993) [Low 
Temp. Phys. {\bf 19}, 670 (1993)]; A. M. Kadigrobov, R. I. Shekhter, and 
M. Jonson, Physica B: Condens. Matter {\bf 218}, 134 (1996).
\bibitem{23}Yu. N. Chiang and O. G. Shevchenko, Fiz. Nizk. Temp. {\bf 25}, 
432 (1999) [Low Temp. Phys. {\bf 25}, 314 (1999)].
\bibitem{24}Yu. V. Nazarov and T. H. Stoof, Phys. Rev. Lett. {\bf 76}, 823 
(1996).
\bibitem{25}L. G. Aslamazov, A. I. Larkin, and Yu. N. Ovchinnikov, Zh. Eksp. 
Teor. Fiz. {\bf 55}, 323 (1968) [Sov. Phys. -- JETP {\bf 28}, (1968)].
\bibitem{26}A. I. Svidzinskii, {\it Prostranstvenno--neodnorodnye zadachi 
teorii sverkhprovodimosti}, Moscow; Nauka (1982) [{\it Spatial--nonhomogeneous 
problems in the theory of superconductivity}].
\bibitem{27}H. A. Blom, A. Kadigrobov, A. M. Zagoskin, R. I. Shekhter, and 
M. Jonson, Phys. Rev. B {\bf 57}, 9995 (1998-II).
\bibitem{28}B. Z. Spivak and D. E. Khmel'nitskii, Pis'ma Zh. Eksp. Teor. Fiz. 
{\bf 35}, 334 (1982) [JETP Lett. {\bf 35}, 413 (1982)].
\bibitem{29}I. O. Kulik, Zh. Eksp. Teor. Fiz. {\bf 57}, 1745 (1969) [Sov. 
Phys. -- JETP {\bf 30}, 944 (1970)].
\end{thebibliography}
\end{document}